\documentstyle[aps,epsf,multicol]{revtex}

\title{Burgers turbulence with pressure}
\author{S.~A.~Boldyrev \\
\em{Princeton University, P.O.Box 451, Princeton, NJ 
08543}}
\date{July 31, 1997}
\begin{document}
\input psfig.sty
\maketitle
\begin{abstract}

The randomly driven Burgers equation with pressure is considered 
as a 1D model of 
strong turbulence of compressible fluid. It is shown that 
infinitely small 
pressure provides a finite effect on the velocity and density 
statistics and this
case therefore is qualitatively different from turbulence without 
pressure. We establish the corresponding operator product expansion 
and predict 
the intermittent velocity-difference and mass-difference PDFs.  We 
then apply 
the developed methods to the statistics of a passive scalar advected 
by the 
Burgers field.\\
~\\
\noindent PACS Number(s): 47.27.Gs, 03.40.Kf, 52.35.Ra.
\end{abstract}

\begin{multicols}{2} 

The Burgers equation with a random external force is considered to be 
the first exactly
solvable model of 1D turbulence and has been extensively studied 
in recent years
\cite{Polyakov,Y-Ch,Boldyrev,F-D,Bouchaud,Sinai,G-M,Balkovsky,Gotoh,Ivashkevich}. 
Though rather simplified, this model can serve as a test model for 
some general 
ideas within the theory of strong turbulence. Recently, methods of 
quantum field 
theory were applied to this problem by A.~Polyakov
\cite{Polyakov}, which enabled a qualitative explanation of the numerical 
findings
by A.~Chekhlov and V.~Yakhot \cite{Y-Ch}. In \cite{Boldyrev} it was shown 
that the
Polyakov approach allows, in fact, to obtain the quantitatively correct 
results. 
We believe that the operator product expansion, introduced in
\cite{Polyakov} to take into account the viscous term, is an adequate 
language to
treat compressible turbulence (even in higher dimensions, where shock
structures and associated local dissipation persist). In the present 
paper 
we extend these ideas to Burgers turbulence with pressure, a more 
realistic 
model of 1D turbulence.  

The basic equations we will study are the following:
\begin{eqnarray}
 & u_t + uu_x = \nu u_{xx} - c^2 \rho^{\gamma-2} \rho_x + f \,\,\,, 
\nonumber \\
 & \rho_t + (\rho u)_x = 0 \,\,\,, \label{Eq.1}
\end{eqnarray}
\noindent where we assume pressure of the form $p=c^2 \rho^{\gamma}$. It 
is the simplest
model where we take into account the back reaction of the density field 
to the
velocity dynamics. For~$\gamma=1$, $c$~coincides with the sound velocity. 
The force is chosen to be Gaussian with zero mean
and white in time variance:
\begin{eqnarray}
\langle f(x, t)f(x^{\prime}, t^{\prime}) \rangle = \delta(t-t^{\prime})
\kappa(x-x^{\prime})\,\,,
\label{Eq.0}
\end{eqnarray}

\noindent where the $\kappa$~function is concentrated at some large 
scale~$L$. We
assume that steady states for both velocity and velocity difference 
exist; for
this we can require, for example, that periodic boundary conditions 
on a scale much
larger that $L$ are imposed, and that the zero harmonic in the 
$\kappa$~function is
absent.

The considered model has many applications for fluid and plasma 
problems; instead 
of kinematic pressure it can contain magnetic pressure and thus 
describe MHD 
turbulence, etc. The resulting equations for the 1D case are quite 
similar to 
system (\ref{Eq.1}). Their derivation and underlying physics can 
be found 
in \cite{Thomas,Yanase}. These equations with white in time and 
space random forcing 
were used to analyze the MHD turbulence in  \cite{F-D} .

In this paper we appeal to the results obtained for Burgers 
turbulence
without pressure in~\cite{Polyakov,Y-Ch,Boldyrev}. In particular, 
we are
interested in the velocity-difference probability density 
function~(PDF), where both
velocities are taken at the same time. The physical picture, 
presented in
these papers, allows us to consider such a general phenomenon as 
intermittency on a
rigorous basis; it is related to the spontaneous breakdown of the 
Galilean 
invariance of the forced equation and to the algebraic decay of the 
PDFs. We will not
repeat these arguments here, instead, we will concentrate on the 
main ideas
which allow us to include the pressure effects into the picture. 

First, we would like to note that equation~(\ref{Eq.1}) develops
singularities (shocks), and that both dissipation and pressure 
terms play a 
regularizing role. The viscous term regularizes the theory at small 
scales and 
also provides energy dissipation. Besides this, it does not make 
much physical
sense. Strong shocks have the width of the order of the mean free 
path, where the gradient
expansion in~$\lambda_{mfp}{\partial \over \partial x}$ used to 
derive the 
viscous term in~(\ref{Eq.1}) does not hold. In principle, this term 
can be chosen in many different 
ways and the steady state can be sensitive to this choice. The general 
results we
disscuss below will account for such a possibility. The same is true 
for the
pressure term in~(\ref{Eq.1}), the role of which will be clarified 
later.

We start by neglecting the external force and considering the special 
stationary solutions of these equations: shock 
waves or kinks. Due to the Galilean invariance of~(\ref{Eq.1}), we can 
work in the 
system, where the shock front is at rest. The second equation 
gives $\rho = B/u$   
\noindent with some constant~$B$. Let~$u_1$ denote the fluid velocity 
to the left of the
shock and~$u_2$ to the right. Obviously, there can be either~$u_1 > 0$, 
$u_2 >0$
or~$u_1 <0$, $u_2<0$, due to continuity. Assume first that~$u_1 > 0$, 
$u_2 >0$,
then~$B>0$. For simplicity, choose $\gamma=1$. The solution for the 
resulting equation 
has the form:
\begin{eqnarray}
{{(u-u_1)^{u_1}}\over{(u-u_2)^{u_2}}} = \exp{{B(u_1-u_2)}\over{\nu}}  
x\,\,\,. 
\label{Eq.2}
\end{eqnarray}
\noindent where $u_1u_2=c^2$. From this solution one can see that 
$u_1>u_2$ and
$\rho_1<\rho_2$, so we have a kink (``$-\tanh$"-like jump) for the 
velocity and 
an antikink (``$+\tanh$"-like jump) for the 
density. For the other
case, $u_1 <0$, $u_2<0$, we will have  kinks for both velocity and 
density. 

Two different kinds of shocks are thus possible for the flux with 
pressure. For
both of them the velocity has a ``$-\tanh$"-like jumps, as in 
turbulence without pressure, 
while the corresponding density profiles are different: 
both ``$-\tanh$" and ``$+\tanh$" jumps are 
allowed. Physically, this corresponds to shock waves moving in 
opposite
directions. This picture does not change qualitatively for $\gamma>1$.

Now we turn on the force and consider the dynamics of the shocks 
with small 
pressure,~${\langle p \rangle \over
\langle \rho \rangle \langle u^2 \rangle } \ll 1$. 
When the shock is just created or two shocks
with density kink and antikink collide, a spike appears in the density 
field,
a $\delta$-singularity, which is regularized by the pressure. One can 
say that
while the viscosity regularizes the velocity field, the pressure term
regularizes the density field. The presence of pressure
(arbitrarily small) can
then result in a repulsion of two colliding shocks. To describe this 
process we note
that there exists a length, characteristic to such a collision:
\begin{eqnarray}
l_p \sim \left({c^2}/{u^2}\right)^{1\over\gamma-1}\,\,\,. \label{Eq.3}
\end{eqnarray} 

This is a characteristic distance at which the collision takes place. 
It depends on
the local velocity $u$, but for estimates one can take $u\sim u_{rms}$,
$u_{rms}=(L\kappa(0))^{1/3}$. 
For the marginal
case $\gamma=1$, this length vanishes, and the collision occurs at the
dissipative scale. (It is the case when the small pressure does not 
affect the
statistics of velocity and can be considered as a perturbation.) 
Analogously, we introduce the dissipative length, which for the second 
order
dissipative regularization takes the form~$l_{\nu}\sim
{\nu/u}$. We will assume that~$\gamma >1$, and set~$\nu$ and~$c^2$ to be 
small.
More precisely, we will be interested in the field correlators at scales 
much larger 
than the dissipative and pressure ones.

To proceed on the quantitative level, we introduce the following~$Z$ 
function: 
\begin{eqnarray}
Z(\mu,y)=\langle \rho(x+y/2)\rho(x-y/2)e^{\mu[u(x+y/2)-u(x-y/2)]} \rangle
\label{Eq.4}.
\end{eqnarray}

As it was shown in~\cite{Polyakov}, this function satisfies the differential 
master equation, which is closed without dissipative and pressure terms. Our 
goal is
to extend the assumptions on the operator product expansion~(OPE), Galilean 
and scaling
invariance introduced in~\cite{Polyakov} to take into accout these terms. 
Nevertheless, we would like to stress that all that will follow is a 
self-consistent 
conjecture. 

If the 
external force has a correlator~$\kappa(y)=1-y^2$, the master equation takes 
the form:
\begin{eqnarray}
{\partial^2 Z\over\partial \mu \partial y} - \mu^2 y^2 Z = D\,\,\,.
\label{Eq.5}
\end{eqnarray}

Here $D$ is the anomaly term coming from both dissipation and pressure in 
the limit
of small~$\nu $ and~$c^2$. Without
further assumptions, this term can not be rewritten in terms of the~$Z$ 
function.  
To find its structure we consider the products of operators, entering~$D$:
\begin{eqnarray}
\nu u_{xx} \rho e^{\lambda u(x)}
\label{Anomaly1}
-c^2 \rho^{\gamma-1} \rho_x e^{\lambda u(x)}\,\,\,.
\end{eqnarray}

The idea now is to consider both of them using the OPE. To do this, we first 
note
that the OPE should be invariant under the rescaling of~$\rho$: 
$\rho\rightarrow \alpha
\rho$. This rescaling leads simply to the rescaling of the cutoff 
parameter~$c^2$, and 
therefore should not affect the statistics at large scales. We have 
the following UV finite operators at our disposal:~$e^{\lambda u}$, $u_x$,
$\rho$. The only Galilean invariant and finite combination satisfying the 
rescaling
condition is~$\rho e^{\lambda u}$. Therefore, we assume the following fusion 
rules for
$\nu \rightarrow 0$ and $c^2 \rightarrow 0$:
\begin{eqnarray}
D=aZ\,\,\,. \label{Anomaly3}
\end{eqnarray}

If the universal  
scaling invariant solution of~(\ref{Eq.5}) exists, its general form is 
given by:
\begin{eqnarray}
Z=F(y)Z_0(\mu y),\,\, F(y)=y^{-2\delta}\,\,, \label{Eq.6}
\end{eqnarray} 

\noindent where $\delta$ is some scaling exponent. Obviously, it should
depend on the order in which we take the non-commuting limits 
$\nu \rightarrow 0$ 
and $c^2\rightarrow 0$, since these two limits correspond to different 
dynamical
pictures. Substituting this into~(\ref{Eq.5}) 
we obtain the following Eq. for~$Z_0$:
\begin{eqnarray}
{\partial^2 Z_0\over\partial \mu \partial y} - \mu^2 y^2 Z_0 = aZ_0 +
{2\delta\over\mu}{\partial\over\partial y}Z_0.
\label{Eq.7}
\end{eqnarray}
\noindent which formally coincides with the Polyakov equation for the 
velocity-difference PDFs of Burgers turbulence \cite{Polyakov}.
Since the two-point density correlator exists,  $Z_0(\mu y)$ tends to a
constant as $\mu \rightarrow 0$. Therefore we should look for the solution 
for~$Z_0$ 
among the family of positive, finite, and normalizable solutions 
of~(\ref{Eq.7}). Such
solutions exist for one parametric family of parameters~$a$ and~$\delta$ 
$(\delta >
0)$, and were obtained in~\cite{Boldyrev}. 

With our present understanding we can not find the anomaly~$a$  from 
the general
consideration. In principle, it can depend on the parameter~$\gamma$ in 
the pressure 
term, exactly in the same manner as the form of the viscous 
regularization can affect 
the steady state. We encounter the same problem in the equation for the 
velocity-difference PDF. 
The operator product $(\nu  u_{xx} - c^2 \rho^{\gamma-2}\rho_x)e^{\lambda
u}$ entering~(1) should be expandable in $e^{\lambda
 u}$ and $u_x e^{\lambda u}$ by the same rescaling arguments as 
in~(\ref{Anomaly3}),
though the coefficients of the expansions are undetermined. 

Nevertheless, simple
arguments allow us to predict a rather peculiar structure of the 
velocity-difference PDF, in particular its left tail, which is 
determined by shocks
\cite{Polyakov,Y-Ch}. This tail has an algebraic decay for $\Delta 
u\rightarrow
-\infty$, and is defined by the possibility of having a shock between 
two points $x-y/2$
and $x+y/2$. To describe it we first note that the dissipative and 
pressure lengths, as the functions 
of local velocity, are equal at some cross-over point~$u_*$. 
If $y/L \ll u_*/u_{rms} \ll 1$, there will be two different decay regimes: 
for $\Delta u \gg u_*$, and for $\Delta u \ll u_*$. For example, for 
the quadratic 
viscous regularization we have $u_*=\nu^{(\gamma-1)/(\gamma-3)}
/c^{2/(\gamma-3)}$. 
For $\gamma<3$, the $\Delta u\gg u_*$ tail will be the same as 
without pressure \cite{Polyakov,Y-Ch,Boldyrev}, while the part 
$\Delta u \ll u_*$ 
will be determined by the pressure.
Moreover, this latter part will decay faster than the former one, 
since the 
pressure term splits large shocks into smaller ones. This function 
for the regular
external force $\kappa(y)=1-y^2$ and for $\gamma < 3$ is
schematically presented on~Fig.~1.

{\columnwidth=3in

\begin {figure} [tbp]
\centerline {\psfig{file=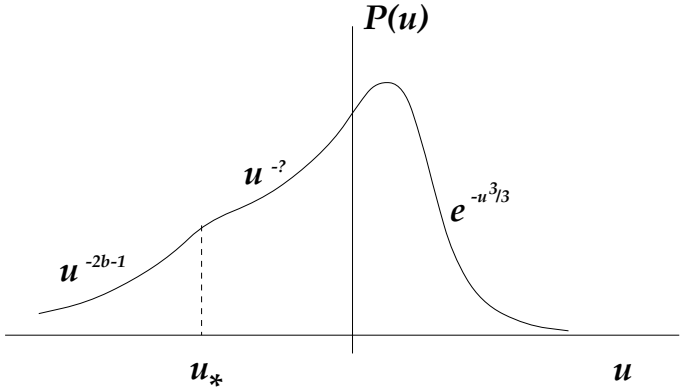,width=9cm}}
\caption{ Velocity-difference PDF for turbulence with 
pressure ($\gamma < 3$). 
We refer to $\Delta u$ simply as $u$.}
\label{fig1}
\end{figure}

}

To complete the picture we now need to describe the statistics of the 
density field.
We will describe it by introducing the probability of having 
mass~$\Delta m$ between two points at the distance $y \ll L$. This PDF is 
defined for
$\Delta m>0$. Presence of pressure leads to large mass correlation at large 
distances, and to the existence of the universal part of the 
mass-difference PDF. 
Due to the scaling invariance the general form of the universal part is 
expected to 
be $P(\Delta m, y)=p(\Delta m/y^{\eta})/y^{\eta}$, where the exponent $\eta$ 
depends on $\gamma$. 
Consistence with (\ref{Eq.6}) fixes this exponent to be 
$\eta = 1 - \delta(\gamma)$. 
We suggest that the universal part of this PDF has an algebraic tail
\begin{eqnarray}
P({\Delta m}) \sim {1\over (\Delta m)^{\theta}}, \,\,\, 
{\bar\rho}y\ll\Delta m \ll
{\bar \rho}L_0.
\label{massPDF}
\end{eqnarray}

\noindent where the $\theta$ exponent is determined by the 
pressure regularization. The scaling invariance breaks for 
large~$\Delta m$, 
since the total mass is conserved ($L_0$ is
the dimension of the system), and also for small~$\Delta m$, 
due to the existence 
of~$\bar \rho$.

This physical picture gives a rather simple prediction for the 
mass-difference structure
functions $\langle (\Delta m)^n \rangle $. Beginning with some~$n$, 
these functions
are non-universal. Their non-universal behavior corresponds to large mass
accumulations between two points, and therefore coincides with the 
behaviour without
pressure, when all the mass is accumulated on shocks. The same
non-universality should show up if we fix~$n$ and 
change~$\gamma$. Numerical data for the 
$\langle \rho(0)\rho(y)\rangle$ correlator do reveal the existence 
of such a nonuniversality, and 
show that the critical $\gamma $ for this case is $\gamma_0 \approx 2$. 
For~$\gamma < 2$, the second order density statistics do not 
feel the pressure
anomaly and $F(y)\equiv\langle \rho(0)\rho(y)\rangle$ approaches 
$F(y)\sim y^{-2}$ as
we decrease $c^2$, as it was without 
pressure\footnote{This result has a simple qualitative explanation. 
One can show that 
the continuity equation with the velocity taken
in the form of the Burgers shock, $u=-U_0\tanh(xU_0/\nu)$, has a 
quasistationary 
solution
$\rho(x, t)\sim 1/\vert x \vert$ for $\frac{\nu^2}{U_0^2}  \gg x^2 \gg 
\frac{\nu^2}{U_0^2}e^{-2tU_0/\nu}$.
Therefore, 
$$\langle \rho(0)\rho(a)\rangle \sim \frac{1}{a}\int^{a-\epsilon}_0
\frac{1}{\vert x-a\vert}\frac{1}{\vert x+a\vert}dx \sim 
\ln (a)/a^2\,\,,$$ 
\noindent where 
we average
over the shock position $x$ between the two points~$-a$ and~$a$. 
The contribution comes from small enough shocks, with $U_0\sim \nu/a$.
We did not
distinguish $\ln$ corrections in the numerical simulations.}. 
Since large mass accumulations occur at the moments of collisions of 
shocks, one can 
say that the shock gas becomes strongly collisional for $\gamma < 2$. 
For $\gamma >
2$, the main contribution to the density correlator comes from the
universal part of the PDF (\ref{massPDF}), and the corresponding 
exponent depends on
the pressure regularization. Our numerical simulations seem to be in 
good agreement
with this picture. We have considered the limit $y \gg l_p \gg \l_{\nu}$, 
and checked 
that for $\gamma<2$, the exponent $2\delta$ tends to 
 $2$ as we decrease $l_p$; the same results as we observed without 
pressure. 
For $2 < \gamma < 3$ the exponent $2\delta$ tends to a limit 
which does depend on $\gamma$. It is interesting that this dependence 
is linear and 
is rather close to $2\delta = (\gamma + 1)/2$.

We emphasize once again that this general consideration does not 
enable us to find
exponents of the tails of velocity-difference and mass-difference 
PDFs, determined 
by the pressure; the exact methods should be
developed for this purpose \cite{Sinai,G-M,Balkovsky,Gotoh,Ivashkevich}. 
Nevetheless, 
the presented picture is rather suggestive and
allows for numerical checks. For example, it predicts the following 
conditional mean
for large negative~$\Delta u$:

\begin{eqnarray}
\langle \rho(0)\rho(y)\vert \Delta u  \rangle \sim y^{-2b}\left({\Delta
u}\right)^{2b-2\delta}\,\,,
\label{Conditional mean 1}
\end{eqnarray}

\noindent where~$b$ is the anomaly for the velocity-difference PDF
\cite{Polyakov,Boldyrev}. The detailed
numerical simulations will be carried out later and presented elsewhere.

In conclusion we would like to briefly discuss  
another important application, a passive scalar advected by the 
Burgers velocity
field. It can be density, concentration or charge of admixture 
particles, temperature 
field or weak magnetic field in a turbulent flow, etc. Using the 
methods for
obtaining the two point
density correlator one can immediately predict the two point correlator 
for the
passive scalar. Indeed, if we neglect the driving and dissipative 
terms, the 
derivative of the passive scalar field~$T_x$ is governed by the 
continuity 
equation:
\begin{eqnarray}
T_{xt} + (uT_x)_x = 0 \,\,\,.
\end{eqnarray}

\noindent Admixture density and charge correspond to the field~$T_x$, 
admixture 
concentration, temperature field and magnetic field correspond to 
the field~$T$.
 Though the problems are not exactly identical (the $T$~field does not
affect the velocity dynamics and $T_x$~should not, in general, be 
strictly
positive), the scale invariant solution and the anomalies should be 
written in the
same form. The basic equation for the passive scalar advection is:
\begin{eqnarray}
T_t + uT_x = \nu_1 T_{xx} + f_1\,\,\,, \label{Eq.8}
\end{eqnarray}   

\noindent where $f_1$ is white in time Gaussian with zero mean 
and the 
correlator $\kappa_1(y)=1-y^2$.
This correlator is assumed to be concentrated at some large~$L_T$. 
By analogy with the
density, one
can write the closed equation for the following $Z$~function: 
$Z_2(y, \mu)=\langle
T_x(x+y/2)T_x(x-y/2)e^{\mu \Delta u}\rangle$. Considering the anomaly:
$$
\nu u_{xx} T_x e^{\lambda u} + \nu_1 T_{xxx} e^{\lambda u}\,\,\,,
$$

\noindent we note that the  translation  symmetry of 
equation (\ref{Eq.8}) 
($T\rightarrow T+a$) is spontaneously broken by the external force 
($T_{rms}=(\kappa_1(0)L_T)^{1/3}$), but restores
for large $L_T$. We also assume the scaling symmetry of the 
fluctuations. 
The presence of such symmetries means that the OPE should be 
proportional to $T_x$. Therefore, the only possible finite and 
Galilean 
invariant form 
is: $T_x e^{\lambda u}$. Then, we have the
following equation for the $Z_2$ function:
\begin{eqnarray}
{\partial^2 Z_2\over\partial \mu \partial y} - \mu^2 y^2 Z_2 
= a Z_2 -
\kappa_1^{\prime \prime}(0)\tilde Z_2\,\,\,,
\label{Eq.9}
\end{eqnarray}

\noindent where $\tilde Z_2$ is the $Z$~function for the velocity 
difference, and is
known.  Now we assume
that for $y \rightarrow 0$ there exists the scaling solution, 
$Z_2=y^{-2\delta}Z_0$.
This can only be true if for small $y$ we can neglect the last term in
Eq. (\ref{Eq.9}): $a Z_2 \gg \kappa_1(0)\tilde Z_2$, i.e. the solution 
should be
dominated by the zero mode of (\ref{Eq.9}). It is important that
this inequality hold uniformly for any $\mu$. The 
following consideration shows that this can be satisfied only in 
the case $Z_0 \equiv \tilde Z_2$. 
Consider the
asymptotics of the velocity-difference PDF (which is the Laplace 
transform of $\tilde
Z_2$):
\begin{eqnarray}
& w(u)\sim u^{-2b - 1} \,,\,\, u \rightarrow -\infty \nonumber\\ 
& w(u)\sim u^{2b-1}e^{-u^3/3}\,,\,\, u \rightarrow +\infty
\end{eqnarray}

\noindent One can obtain the asymptotics of the Laplace transform 
of~$Z_0$ by simply
changing $b
\rightarrow \delta$ in these expressions. Comparison of these 
asymptotics shows that the inequality holds 
uniformly with respect to~$u$ only when~$\delta=b$.  In other 
words, the only zero mode 
surviving in the limit $y/L\ll 1$ is the mode with~$\delta=b$.

One can therefore predict for the second order structure function:

\begin{eqnarray}
\langle (T(0)-T(y))^2 \rangle \sim y^{2-2b} \,,\,\,\, y
\rightarrow 0 \,\,.
\end{eqnarray}

\noindent For each $\alpha$ in the short-distance 
expansion $\kappa(y)=1-y^{\alpha}$ some particular
solution $b=b(\alpha)$ realizes. For the case of a regular 
external force,
$\alpha=2$, it is argued in \cite{Boldyrev}  that $b=1/2$, 
and  we obtain 
$\langle (T(0)-T(y))^2 \rangle \sim y\ln(y)$.\\
{}\\

I am grateful to A.~Polyakov and V.~Yakhot for useful 
discussions and comments. 
I would also like to thank M.~Chertkov, T.~Gotoh, V.~Gurarie, 
E.~Ivashkevich, 
and D.~Uzdensky for important discussions on both the physics 
and the numerics of 
the problem, J.~Krommes and A.~Schekochikhin for their help 
in numerical 
simulations, and T.~Munsat for valuable remarks on the 
style of the paper.

This work was supported by U.S.D.o.E. Contract 
No. DE-AC02-76-CHO-3073 and
also by ONR/DARPA grant No. N00014-92-J-1796.

\end{multicols}

\end{document}